\documentstyle[12pt,graphicx]{article}

\def\beq{\begin{equation}}
\def\eeq{\end{equation}}
\def\beqa{\begin{eqnarray}}
\def\eeqa{\end{eqnarray}}
\newcommand{\gsim}{ \mathop{}_{\textstyle \sim}^{\textstyle >} }
\newcommand{\lsim}{ \mathop{}_{\textstyle \sim}^{\textstyle <} }
\newcommand{\vev}[1]{ \left\langle {#1} \right\rangle }

\newcommand{\EV}{ {\rm eV} }
\newcommand{\KEV}{ {\rm keV} }
\newcommand{\MEV}{ {\rm MeV} }
\newcommand{\GEV}{ {\rm GeV} }
\newcommand{\TEV}{ {\rm TeV} }

\begin{document}
\baselineskip 0.6cm

\def\simgt{\mathrel{\lower2.5pt\vbox{\lineskip=0pt\baselineskip=0pt
           \hbox{$>$}\hbox{$\sim$}}}}
\def\simlt{\mathrel{\lower2.5pt\vbox{\lineskip=0pt\baselineskip=0pt
           \hbox{$<$}\hbox{$\sim$}}}}

\begin{titlepage}

\begin{flushright}
UCB-PTH-04/26 \\
LBNL-55487 
\end{flushright}

\vskip 2.0cm

\begin{center}

{\Large \bf 
Simultaneous Solutions of the Strong CP and $\mu$ Problems
}

\vskip 1.0cm

{\large
Brian Feldstein, Lawrence J. Hall, and Taizan Watari
}

\vskip 0.4cm

{\it Department of Physics, University of California,
                Berkeley, CA 94720} \\
{\it Theoretical Physics Group, Lawrence Berkeley National Laboratory,
                Berkeley, CA 94720}

\vskip 1.2cm

\abstract{
The $\mu$ parameter of the supersymmetric standard model is replaced 
by $\lambda S$, where $S$ is a singlet chiral superfield, introducing 
a Peccei--Quinn symmetry into the theory.
Dynamics at the electroweak scale {\em naturally} solves both 
the strong CP and $\mu$ problems as long as $\lambda$ is of order 
$\sqrt{M_Z /M_{\rm pl}}$ or smaller, and yet this theory has 
the same number of relevant parameters as the supersymmetric standard 
model.
The theory will be tested at colliders: the $\mu$ parameter is 
predicted and there are long-lived superpartners that decay to 
gravitinos or axinos at separated vertices. 
To avoid too much saxion cold dark matter, a large amount of entropy 
must be produced after the electroweak phase transition. 
If this is accomplished by decays of a massive particle, the reheat 
temperature should be no more than a GeV, strongly constraining 
baryogenesis. 
Cold dark matter may be composed of both axions, probed by direct 
detection, and saxions, probed by a soft X-ray background arising from 
decays to $\gamma \gamma$. 
There are two known possibilities for avoiding problematic axion 
domain walls: the introduction of new colored fermions or the assumption 
that the Peccei--Quinn symmetry was already broken during inflation.
In the first case, in our theory the colored particles are expected 
to be at the weak scale, while in the second case 
it implies a good chance of discovering isocurvature perturbations 
in the CMB radiation and a relatively low Hubble parameter 
during inflation.
}

\end{center}
\end{titlepage}

\section{Introduction}

A spontaneously broken global symmetry remains an attractive 
solution to the strong CP problem \cite{PQ}. 
The strong CP parameter $\bar{\theta}$ is canceled 
by the dynamical relaxation of the resulting pseudo-Goldstone boson, 
the axion \cite{WW}.
A global symmetry with a QCD anomaly, U(1)$_{\rm PQ}$, can be 
implemented in one of the simplest extensions of the standard 
model---models with two Higgs doublets $h_{1,2}$ \cite{PQ}.
In order that $f_a$ be much larger than the weak scale, $\tilde{m}$, 
the primary breaking of U(1)$_{\rm PQ}$ must come from an electroweak
singlet scalar, $s$. As in the DFSZ invisible axion models
\cite{DFSZ}, these scalars will have U(1)$_{\rm PQ}$-preserving 
interactions such as $s h_1 h_2$ or $s^2 h_1 h_2$, but not 
$s^* h_1 h_2$ or $s^{*2} h_1 h_2$. 
This extension of the standard model fits well with supersymmetry (SUSY), 
except for an immediate question of why the masses of the two Higgs 
doublets are much smaller than the scale $f_a$. 
This corresponds to the $\mu$ problem in the supersymmetric standard 
model.

In this paper we point out an extremely simple model that
simultaneously solves both the strong CP and $\mu$ problems without 
severe fine-tuning.  
Ours is certainly not the first such theory (i.e., SUSY DFSZ axion 
models) \cite{SUSY-DFSZ,SUSY-DFSZ2,SUSY-DFSZ3}, 
but it is very simple and has important 
consequences for signals at accelerators and in cosmology.
In particular, the gravitino and axino (fermionic SUSY partner of 
the axion) are much lighter than the weak scale, so that the superpartners 
produced at hadron colliders end up decaying to either gravitinos or 
axinos possibly with separated vertices. 
The saxion, the scalar SUSY  partner of the axion, is also lighter than the 
weak scale.
All SUSY particles around the weak scale are unstable, 
but cold dark matter may be composed of axions,\footnote{ 
It is remarkable that the cosmological dynamical relaxation 
of the axion field during the QCD era can lead to 
the observed amount of dark matter in cold axions, providing 
the symmetry breaking scale is $f_a \approx 10^{11}$ GeV, 
one order of magnitude above the lower bound set by 
the SN 1987A constraint. 
This value for $f_a$ required for the dark matter may be obtained 
as the geometric mean of the supersymmetry (SUSY) breaking scale 
and the Planck scale \cite{SUSY-DFSZ}.} and possibly 
saxions. The cosmological saxions lead to astrophysical X-ray signals.

The model is presented in section \ref{sec:model}. 
Limits on the SUSY-breaking scale, and hence on the gravitino mass, 
are discussed in section \ref{sec:gravitino}. 
Section \ref{sec:thermal} is devoted to the thermal history of 
the model, including late-time entropy production required 
from saxion evolution. The final section contains a summary 
of the predictions of our theory, along with remarks on the difference 
of our theory from those in \cite{SUSY-DFSZ}.

\section{Model}
\label{sec:model}

We consider a supersymmetric theory at the electroweak scale with a
superpotential 
\beq
W = \lambda S H_1 H_2
\label{eq:model}
\eeq
together with Yukawa interactions of the quarks and leptons to the
Higgs doublets $H_{1,2}$.
This is certainly a very simple theory: it is the minimal supersymmetric
standard model (MSSM) with $\mu$ replaced by $\lambda S$, where $S$ is a gauge
singlet superfield, and the soft parameter $\mu B$ replaced by $A 
\lambda s$. Alternatively it can be viewed as the next-to-MSSM 
without the $\kappa S^3$ interaction. 
There is a U(1)$_{\rm PQ}$ symmetry that is spontaneously broken by 
$\vev{s}$, leading to a Goldstone boson. 
Therefore, $\vev{s}$ must be larger than about $10^{10}$ GeV
to avoid laboratory and astrophysical constraints. 
One immediate problem is to obtain an appropriate scalar potential for 
$s$; $\vev{s}$ should be non-zero, but should not be infinite.
The superpotential (\ref{eq:model}) alone neither destabilizes nor 
stabilizes the flat direction $s$.

This theory has been studied before as a solution to the strong CP 
problem \cite{MN}, and a stable vacuum with finite $\vev{s}$ was found 
in the analysis of the scalar potential involving SUSY breaking.
However, the authors of \cite{MN} believed 
that a fine-tune was necessary to obtain a stable vacuum, 
and that an effective $\mu_{\rm eff} \equiv \lambda \vev{s}$ 
parameter of order weak scale, $\tilde{m}$, is obtained 
only as a result of another accidental cancellation between 
$\vev{s} \gg \tilde{m}$ and $\lambda \ll 1$.
The $\mu$ problem was not solved.

On the other hand, it is known \cite{CP,HW} that the theory 
with (\ref{eq:model}) has a natural solution to the $\mu$ problem,  
provided that the soft mass term for $s$ is sufficiently small 
\beq
| m_S^2 | < \lambda^2 \tilde{m}^2
\label{eq:ms}
\eeq
for some dynamical reason (see section \ref{sec:gravitino}).  
Once the two Higgs doublets $H_{1,2}$ acquire vevs $v_1 = \cos \beta
\; v$ and $v_2 = \sin \beta \; v$, the potential for $s$ becomes
\beq
V(s) = - (A \lambda v_1 v_2 s + h.c.) + \lambda^2 (v_1^2 + v_2^2) s^*s,
\label{eq:Vs}
\eeq
leading to a stable vacuum with $\vev{s} \sim \lambda^{-1} \tilde{m}$ 
and generating an effective $\mu$ parameter 
\begin{equation}
\mu_{\rm eff} \equiv \lambda s = A \cos \beta \sin \beta 
\label{eq:mu} 
\end{equation}
of order $\tilde{m}$ for {\it any} value of $\lambda$, providing a
very elegant solution to the $\mu$ problem. 
The crucial point is that the scale of $\mu$ is set by $A$, not by 
$\vev{s}$. Thus, our observation is that the theory described by 
(\ref{eq:model}) provides a simple simultaneous solution 
to the strong CP and $\mu$ problems. 

Of course, to satisfy astrophysical limits 
$f_a \approx \vev{s} \gg \tilde{m}$, 
the coupling $\lambda$ must be very small. 
We will demonstrate in section \ref{sec:prdct-cncl} that some models 
can yield $\lambda \approx \sqrt{\tilde{m}/M_P}$, so that the axion 
decay constant $f_a$ lies around $10^{11}$ GeV, naturally giving 
rise to axions as cold dark matter. 

The spectrum of this theory consists of states in the chiral multiplet 
$S$, in addition to those of the MSSM \cite{CP,HW}. 
The saxion mass, $m_s$, is of order $\lambda \tilde{m} \sim 100 \; \EV 
\times (\lambda/10^{-9})$, and the axino mass is of order\footnote{
The light axino is due to a see-saw mechanism: the fermion component 
of the $S$ multiplet has a mass of order 
$(\lambda \vev{h})^2/(\lambda \vev{s}) \sim \lambda^2 \tilde{m}$. 
This can also be understood in terms of symmetries.
Since $S$ is the only multiplet relevant to the axion, 
the axino mass must be Majorana. 
The mass should be proportional to $\lambda^2$ because the 
superpotential (\ref{eq:model}) has a spurious symmetry
under which phases of $S$ and $\lambda$ are rotated in the 
opposite directions.}\raisebox{4pt}{$\!,$}\footnote{When 
the Kahler potential has a non-renormalizable term 
$|S^\dagger S|^2 /M^2$, there is another contribution to 
the axino mass of order $(f_a / M)^2 m_{3/2}$.
For $M$ of order the Planck scale, $M_{\rm pl}$, it does not exceed 
$10^{-3}$ eV, due to the upper limits on $m_{3/2}$ and $f_a$ 
obtained in section \ref{sec:gravitino} and \ref{sec:thermal}, 
respectively. Thus, none of the discussion in this article 
is changed.} 
$\lambda^2 \tilde{m} \sim 10^{-7} \; \EV (\lambda/10^{-9})^2$: 
these very small masses are a distinctive feature of our theory.
There is no stable WIMP dark matter candidate at the TeV scale. 
The axino is the LSP, but is so light that it will not contribute 
to cold dark matter---axions and saxions are the candidates 
for dark matter.

Various astrophysical processes, that led to the constraints 
on the axion, also provide phenomenological limits 
on the light saxion field.\footnote{Production of axinos is 
sufficiently suppressed as $R$ parity requires that they must 
be produced in pairs.} The saxion is light enough to be emitted from 
the interior of horizontal branch stars, and energy loss from saxion 
emission, which would shorten the lifetime of helium-burning stars, 
sets a limit on the Yukawa coupling of the saxion to the electron. 
The emission is dominated by the bremsstrahlung-like process 
$e^- + {}^4 {\rm He} \rightarrow {}^4 {\rm He}+ e^- + {\rm saxion}$, 
and the constraint on the Yukawa coupling is given by 
\cite{Raffelt-book}
\begin{equation}
 \frac{1}{4 \pi}\left(\frac{m_e}{f_a} \sin^2 \beta \right)^2 
    \lsim 1.4 \times 10^{-29}, 
\end{equation}
or equivalently, 
\begin{equation}
 f_a  \gsim 4  \sin^2 \beta \times 10^{10} \;\GEV. 
\end{equation}
Saxions are also emitted from SN 1987A, carrying energy 
away from the supernova.
The energy loss rate through the saxion turns out to be roughly 
the same as that through the axion \cite{Raffelt-book},
so that the astrophysical limit on $f_a$ is  
a little stronger than in conventional axion models.
These more stringent bounds apparently indicate $f_a$ close to 
$10^{11}$ GeV 
so that axions necessarily contribute a significant fraction 
of the dark matter, but this conclusion requires further scrutiny 
since our theory requires entropy dilution of saxion field 
oscillations, as discussed in section \ref{sec:thermal}.

\section{A Light Gravitino}
\label{sec:gravitino}

The stable vacuum of our theory crucially relies on an assumption 
$|m_S^2| < \lambda^2 \tilde{m}^2$. There are models of mediation 
of SUSY breaking with vanishing $m_S^2$, but radiative corrections 
to $m_S^2$ through the interaction (\ref{eq:model}) are also of order 
$\lambda^2 \tilde{m}^2$. Thus, one should expect 
either i) an accidental cancellation between tree-level and one-loop
contributions, or ii) the SUSY breaking is mediated at a low energy 
scale, so that the one-loop correction is sufficiently small.
Let us briefly see how low the mediation scale should be.

The tree-level potential of the CP-even scalars shows that 
the smallest eigenvalue of the mass-squared matrix is positive 
when \cite{CP}
\begin{equation}
 \left| \xi \equiv \frac{m_S^2}{\lambda^2 (v_1^2 + v_2^2)} \right| 
    \lsim \frac{M_Z^2}{M_A^2}.
\label{eq:vac-sta}
\end{equation}
We roughly\footnote{Detailed analysis using $\tan \beta > 2.5$ 
and $\mu > 120$ GeV leads to $-0.15 < \xi < 0.12.$  For the effects 
of 1-loop corrections to the scalar potential, see \cite{MNZ,MN}.} 
take this limit to be $|\xi| \lsim 0.2$.
Since the renormalization-group equation for $m^2_S$ is given by 
\begin{equation}
 \frac{\partial m_S^2(\mu)}{\partial \log{ \mu}} = 
   - 2 \frac{\lambda^2}{8 \pi^2} (m_1^2 + m_2^2 +  A^2 + m_S^2),
\end{equation} 
the one-loop contribution to $\xi$ is of order 
\begin{equation}
 \xi_{1-{\rm loop}} \approx - \frac{1}{2\pi^2} 
    \ln \left(\frac{M_S}{\lambda s}\right), 
\end{equation}
where SUSY breaking is assumed to be mediated at some energy scale 
$M_S$.
Thus, it follows from the vacuum stability condition 
(\ref{eq:vac-sta}) that to avoid any fine-tuning the 
``messenger scale'' $M_S$ is at most 
one order of magnitude higher than the electroweak scale. 
Models with such a low messenger scale are found in \cite{GNS}.
For larger values of the messenger scale, the amount of fine-tuning
increases logarithmically. For example, for
 gauge mediated SUSY breaking models with a messenger scale 
of $10^{3}$ TeV, fine-tuning of order $1/10$ 
is required between $\xi_{\rm tree}$ and $\xi_{\rm 1-loop}$. 

There is no conflict between the requirement of a low messenger scale
and the large value for $\vev{s}$, which is driven by the soft
operators. 
The soft parameters in the one-loop effective potential are 
renormalized at $\lambda s$, because this is the combination 
that appears in particle masses.
Since we are interested in the scalar potential for $s$ with 
$\lambda s$ of order the weak scale, the soft parameters are evaluated 
at the weak scale, where they are local  
no matter how low the messenger scale is.
The scalar potential of the $s$ field is essentially given by 
physics at the electroweak scale, even though the vev 
$\vev{s} \approx f_a$ is much larger than the electroweak scale.

The fine-tuning argument above favors a low mediation scale, 
but does not directly constrain the fundamental scale of 
local supersymmetry breaking  $\sqrt{F_{\rm SUSY}}$. 
However, in supergravity theories all scalars fields 
typically\footnote{When the Kahler potential has a certain form, 
the ordinary gravity-mediated supersymmetry-breaking masses are absent,
and $m_S^2$ acquires only an anomaly-mediated piece, which is 
of order $(\lambda^2/4\pi) \alpha_L m_{3/2}^2$. In this case, 
although $m_{3/2}$ of order the weak scale is allowed, the 
discussion in the following sections is not modified essentially.
In section \ref{sec:thermal}, the gravitino is no longer light, 
but the entropy production required to dilute the saxion oscillation 
also dilutes the gravitino number density, and there is no 
gravitino problem. It is known that the Affleck--Dine mechanism works 
for baryogenesis \cite{KSY}. In section \ref{sec:prdct-cncl}, 
we still expect separated vertices in colliders, although 
they arise from decays to axinos rather than to gravitinos.}  
acquire a supersymmetry breaking mass, giving a contribution to 
$m_S^2$ of order $[m_{3/2} = F_{\rm SUSY} / \sqrt{3} M_{\rm pl}]^2$, 
where $M_{\rm pl} \simeq 2.4 \times 10^{18}$ GeV is the Planck scale. 
For this contribution to satisfy (\ref{eq:ms}) without any fine-tuning,
the bound on the scale of local supersymmetry breaking is 
\begin{eqnarray}
m_{3/2} & \lsim & 100 \; \EV \times (\lambda /10^{-9}), \nonumber \\
\sqrt{F_{\rm SUSY}} & \lsim & 300 \; \TEV 
 \times (\lambda /10^{-9})^{1/2}.
\label{eq:gravmasslim}
\end{eqnarray}

If this bound is saturated we would normally expect a gravitino
problem: the gravitinos are in thermal equilibrium at the weak scale,
and although they are somewhat diluted by later annihilations, they
still give too much hot dark matter. This would lead to an even
stronger bound on  $\sqrt{F_{\rm SUSY}}$ than given above. 
Although the gravitino is not the LSP, it decays to axion-axino 
with a lifetime longer than the age of the universe, so that the 
gravitino problem is not alleviated.  
However, in section \ref{sec:thermal} we see that entropy production  
after the electroweak phase transition is required to dilute 
saxion oscillations, and this will also dilute the gravitinos.

\section{Thermal History}
\label{sec:thermal}

\subsection{Saxions and Late-Time Entropy Production}
Supersymmetric axion models always involve a saxion field 
with a mass at most of order the SUSY breaking scale, $\tilde{m}$.
Thus, there is a flat direction, and its evolution in the early
universe must be examined carefully.

During inflation the saxion field could 
be zero or large, for example of order the Planck scale, depending 
on its coupling to the inflaton. We begin by supposing that it
is at the origin. In this case it stays at the origin
until the Peccei--Quinn phase transition is triggered by the 
Higgs vev at a temperature of order the electroweak scale, as seen 
from the potential of eq.~(\ref{eq:Vs}). 
Immediately after the phase transition, the saxion field oscillates 
about the minimum of its potential, with an energy density, $V_s$,
of order $\tilde{m}^4$.
If the saxion were to decay rapidly enough, for instance, with a decay rate 
of order $\Gamma \sim \tilde{m}^3/f_a^2$, the field energy would
rapidly convert into radiation giving no problem. 
However, the saxion mass is not of order $\tilde{m}$, 
but $\lambda \tilde{m}$. The lifetime of the saxion is of order 
$\tau \sim 10^2 (10^{-9}/\lambda)^5 \times 10^{10}$ yrs., and 
is much longer than the present age of the universe.
The oscillation of the saxion field, which behaves like matter, overcloses 
the universe. To avoid this we study the dilution of the
saxion field oscillations by large entropy production
after the electroweak and Peccei--Quinn phase transitions.

Let us suppose, for simplicity, that the entropy is produced via the
decays of a massive particle, $X$, which could be the 
inflaton, curvaton or flaton. During and after the electroweak phase
transition the universe is dominated by $X$, and is therefore 
matter-dominated. 
While the Hubble parameter is much larger than the decay rate of the 
$X$ particle, the energy density of $X$, $\rho_X$, scales as 
$\propto 1/a^3$, where $a$ is the scale factor.
Some $X$ particles decay at a time much less than the $X$ lifetime, 
producing entropy. The total energy density of radiation 
$\rho_\gamma \sim T_\gamma^4$ does not scale as $\propto 1/a^4$, 
but rather as $\propto 1/a^{3/2}$, because of the continuous 
entropy supply from the $X$-particle decays.
These $X$ decays to radiation clearly dilute the saxion field 
oscillation energy density. 
This dilution continues until the age of the universe becomes 
comparable to the lifetime of the $X$ particle, when $\rho_\gamma$ 
is also comparable to $\rho_X$ (see e.g. \cite{KT}).
A long $X$ lifetime, and therefore a low value for the reheating
temperature $T_R$, leads to more dilution. 
Any initial thermal saxions are diluted to a negligible level, 
while the current number density of cold saxions in
the saxion field oscillation is given by 
\begin{equation}
 \frac{n_s}{n_\gamma} = c \times 10^{-1} 
     \times \left(\frac{T_R}{1 \GEV}\right)^5 
            \left(\frac{100 \GEV}{T_{\rm PQ}}\right)^4
            \left(\frac{100 \EV}{m_s}\right),
\label{eq:ns}  
\end{equation}
where $T_{\rm PQ}$ is the temperature of the Peccei--Quinn 
phase transition, and the dimensionless coefficient $c$ is given by 
\begin{equation}
 c \sim \left(1+\frac{21}{22}\right) \left(\frac{2}{5}\right)^2 
        \frac{3 \zeta(4)}{\zeta(3)} 
        \left(\frac{g_S(T_R)}{g_S(T_{\rm PQ})}\right)
        \left(\frac{V_s}{\rho_{\rm rad}}\right)_{\rm PQ}.
\label{eq:cofns}
\end{equation}
Here, the last factor $V_s/\rho_\gamma$ is evaluated at the epoch 
of the Peccei--Quinn phase transition, and $g_S(T_{\rm PQ})$ 
and $g_S(T_R)$ are the effective statistical degrees of freedom 
when the temperature is around $T_{\rm PQ}$ and $T_R$, 
respectively. 
The saxion field oscillations contribute to the present energy density
an amount  
\begin{equation}
 \Omega_s h^2 = c' \times 10^{-1} 
     \times \left(\frac{T_R}{1 \GEV}\right)^5 
            \left(\frac{100 \GEV}{T_{\rm PQ}}\right)^4,
\label{eq:TR}
\end{equation}
where
\begin{equation}
 c' \sim \left(1+\frac{21}{22}\right) \left(\frac{2}{5}\right)^2 
        \left[ \Omega_\gamma h^2 \frac{100 \; \EV}{2.73 {\rm K}} 
               \simeq 10.5 \right]
        \left(\frac{g_S(T_R)}{g_S(T_{\rm PQ})}\right)
        \left(\frac{V_s}{\rho_{\rm rad}}\right)_{\rm PQ}.
\end{equation}
The saxion energy density does not depend on the choice of $f_a$. 
$\Omega_{\rm CDM} h^2 \approx 0.1$ requires 
$T_R \lsim 1$~GeV, and when this bound is saturated, the saxion 
is also a significant component of the CDM.\footnote{
To be more precise, the vev's of the Higgs fields are determined 
by a thermal potential, and they change as the temperature falls. 
The minimum of the $s$ field is also changing accordingly. 
Thus, even if the entropy production dilutes the energy of the 
saxion oscillation when most of the $X$ particles have decayed, 
one has to further make sure that such readjustments after the entropy 
production do not release too much oscillation energy for the CDM.
It turns out that the Higgs field values are close enough 
to the vacuum values when the temperature is around 1~GeV or lower, 
and the saxion oscillation due to this late-time readjustments 
is not cosmologically important.}

Such a large entropy production not only dilutes saxions 
to an acceptable level, but also dilutes other species.
The SUSY particles of the MSSM sector annihilate quickly to lighter 
SUSY particles, which eventually decay to the gravitino or axino.
Axino number-changing reactions have already decoupled by the
electroweak scale, and gravitino number-changing reactions freeze-out
well before $T_R$, so that both axinos and gravitinos are
significantly diluted by the entropy production.
Hence, even gravitino masses that saturate the bound of 
(\ref{eq:gravmasslim}) do not lead to an amount of hot 
dark matter in conflict with observation. Furthermore,
gravitinos and axinos provide 
negligible contributions to the effective number 
of neutrino generations during BBN and CMB eras. 

Entropy production from $X$ decays also dilutes the baryon asymmetry, 
by a factor $(T_R/T_B)^5$, where $T_B$ is the temperature at which the
baryon asymmetry is generated. This severe dilution
implies that the baryon asymmetry cannot be created 
above the weak scale. One possibility is that a baryon asymmetry
of order unity is created at the weak scale.
Another is that the observed small asymmetry is created 
in the out-of-equilibrium decays of the $X$ particles; 
see, \cite{DH} as an example, where $R$-parity-violating interactions 
$W = UDD$ are required, and $T_R$ has to be less than 1~GeV.

\subsection{Relic Axion Energy Density}
Let us now turn our attention to the energy density carried by the 
axion field.
There are two significant components: one from axions 
emitted by axionic strings, and the other from the misalignment 
of the initial value of the axion field from that of the true 
potential minimum.

The energy density from the axion phase relaxation is known to 
be \cite{Turner}
\begin{equation}
    \Omega_{\rm mis.} \; h^2 = 
      0.10 \times 10^{\pm 0.4} \; 
      \left(\frac{\Lambda_{\rm QCD}}{200 \; \MEV}\right)^{-0.7}
      \left(\frac{f_a}{10^{11} \; \GEV}\right)^{1.18}, 
\label{eq:misalignA}
\end{equation}
in the absence of the entropy production from $X$ decays. 
This calculation assumes a radiation-dominated background at the epoch 
of relaxation, when the effective mass of the axion $m_{a,{\rm eff}} 
\sim 0.1 m_a \times (\Lambda_{\rm QCD}/T_\gamma)^{3.7}$ becomes 
comparable to the Hubble parameter.
If $T_R$ is around its upper bound $\sim 1$ GeV, then this
assumption is justified.
Given the lower bound on $f_a$ at the end of section \ref{sec:model}, 
which is more stringent than in conventional models, in our theory the 
axion energy density accounts for a significant fraction of CDM, 
and axions are clearly a natural candidate for CDM.
When the decay temperature is well below $1$~GeV, 
massive $X$ particles affect the axion energy density in two ways 
\cite{KMY}. The extra energy density from $X$ particles delays 
the epoch of the relaxation of the axion field, and  
the entropy from $X$ decays dilutes the axion energy density. 
Combining both effects, the axion energy density is given by 
\begin{equation}
 \Omega_{\rm mis.} h^2 \approx 0.1 \times 
     \left(\frac{f_a}{10^{11} \; \GEV}\right)^{1.5}
     \left(\frac{T_R}{1 \; \GEV}\right)^{2.0}, 
\label{eq:misalignB}
\end{equation}
and in order to have axion CDM we require that $f_a$ be raised to 
\begin{equation}
 f_a \sim 10^{11} \; \GEV \left(\frac{1 \; \GEV}{T_R}\right)^{1.3}.
\label{eq:fatd}
\end{equation}
Imposing $T_R \gsim 1$ MeV from big bang nucleosynthesis, 
the upper bound on $f_a$ is $f_a \lsim 10^{15}$ GeV \cite{KMY}.

If $s$ is large during inflation then the axionic strings from the PQ
phase transition are inflated away, and the axion energy density is from 
the misalignment of the initial phase.
However, if $s$ vanishes during inflation, axionic strings are formed 
after the PQ phase transition is triggered at the electroweak scale. 
After further cooling during the $X$ dominated era, 
the energy density in strings reaches a fraction 
$(f_a / M_{\rm pl})^2 \ln (f_a/H)$ of the total energy density. 
When the temperature falls so that $m_{a,{\rm eff}}$ becomes comparable to 
the Hubble parameter, axionic domain walls emerge and the string network 
turns into the boundary of the domain walls. 
This string/domain wall system rapidly disappears by radiating 
axions,\footnote{There must be only one vacuum in the phase 
direction of $s$. Otherwise, such string/domain wall system 
cannot disappear.} and the number density of axions is fixed:
\begin{equation}
 n_{a,{\rm string}} \approx 
 H f_a^2 \Delta, 
\end{equation}
where $\Delta$ ranges from $\ln(f_a/H)$ to unity.
This large uncertainty in $\Delta$ corresponds to the disagreement 
between \cite{Davis} and \cite{HS} about the typical energy 
of axions emitted from the string network.
Since the axion number density from misalignment is also $Hf_a^2$ 
at that epoch, 
the resulting relic density from strings is $\Delta$ times that of 
the misalignment axions, just as in the case of radiation dominance.
Note also that another uncertainty in the relic density 
arises from assuming that the energy density of the string network 
is converted into axion particles when $m_{a,{\rm eff}}$ 
is equal to the Hubble parameter.

In the case that $s$ vanishes after inflation, we have assumed that 
there is only a single vacuum in the direction of the axion field.
However, the model presented in section \ref{sec:model} has three 
vacua, due to three families contributing 
to the U(1)$_{\rm PQ}$[SU(3)$_C$]$^2$ anomaly.
The number of vacua can be reduced to one by introducing extra 
colored particles.
For instance, introducing two vector-like pairs of 
chiral multiplets $\Phi_i({\bf 3})+\Phi^c_i({\bf 3}^*)$ ($i=1,2$), 
with a coupling 
\begin{equation}
 \Delta W = \lambda_i S \Phi_i \Phi^c_i,
\label{eq:extra}
\end{equation}
gives an anomaly coefficient of 1, so that there is a unique vacuum.
If the couplings $\lambda_i$ ($i=1,2$) are much smaller than $\lambda$,  
then these extra particles have already been excluded by data. 
On the other hand, if $\lambda_i$ are larger than $\lambda$,  
the SUSY-breaking masses\footnote{This argument does not apply if
$\lambda_i$ are so large that their masses 
due to (\ref{eq:extra}) are larger than the ``messenger scale''
$M_S$.} of the extra particles contribute 
to $\xi_{\rm 1-loop}$ at order $\lambda_i^2/\lambda^2$, 
requiring excessive fine-tuning. 
Thus, $\lambda_i \approx \lambda$, and, as a consequence, these extra 
colored chiral multiplets are expected 
at the electroweak scale. If these vector-like particles have the same
electric and PQ charges as the up or down-type quarks, then they can decay by
mixing with the known quarks.

\subsection{Early Breaking of PQ Symmetry}
It has been assumed so far in this section that the $s$ field vanishes 
after inflation until later times. 
If $s$ takes a very large value $s_0$ during and after inflation, 
the radial direction of the $s$ field, i.e., the saxion field, 
starts to oscillate when the Hubble parameter is comparable 
to the curvature of its scalar potential. 
The energy of the saxion oscillation has to be diluted in this case, 
as well, by an amount that depends on $s_0$. 
If $s_0$ is of order of the Planck scale, even the entropy production 
from the $X$ decays is not enough.
Indeed, $\rho_s/\rho_X \approx (s_0/M_{\rm pl})^2$ when the $s$ field 
starts to oscillate,\footnote{Here, the gravity-mediated quadratic 
and/or gauge-mediated logarithmic SUSY-breaking potential for $s$ 
is assumed.} and for $X$ decays to provide sufficient dilution for 
saxions, we find that the following condition should be satisfied:
\begin{equation}
s_0  \lsim  s_{0,{\rm max}} \equiv 
10^{-3}  M_{\rm pl} \times \sqrt{\frac{\MEV}{T_R}}. 
\label{eq:smalls0}
\end{equation}

The initial phase of $s_0$ is well-defined and almost homogeneous 
inside the present horizon provided the Hubble parameter during 
the inflation $H_I$ is smaller than $s_0$; 
the phase fluctuation that might be generated during inflation 
is of order $\delta \theta \sim H_I/(2 \pi s_0) < 1$. 
The initial phase is preserved in classical evolution of the $s$ field 
until the QCD phase transition, when the potential in the phase 
direction emerges.
Since the whole universe inside the horizon falls into the 
single vacuum, there is no problem of domain walls even in the 
absence of the extra colored particles. 
The estimate (\ref{eq:misalignA}) or (\ref{eq:misalignB}) 
for $\Omega_{\rm mis.}$ applies to this case, except that 
i) the initial phase of $s_0$ is used instead of the average 
of random phases $\pi/\sqrt{3}$, and 
ii) the normalization of $f_a$ can be different because of 
the different number of vacua in the phase direction \cite{Turner}.
Note that axionic string network is not formed after inflation 
in this case, and the misalignment of the initial phase is the only 
source of the cosmological axions. 

The phase fluctuation $\delta \theta$ leads to isocurvature 
density perturbations in the axions and radiation \cite{iso}, giving 
\begin{equation}
\left( \frac{\delta T}{T} \right)_{\rm isocurv} \approx 
 10^3 \frac{H_I}{2\pi M_{\rm pl}}
 \sqrt{\frac{T_R}{\MEV}} \frac{s_{0,{\rm max}}}{s_0}.
\end{equation}
Since the current CMB data is consistent with purely adiabatic 
density perturbation, this implies $H_I \lsim 10^{11} \; \GEV  
\sqrt{\frac{\MEV}{T_R}} \frac{s_0}{s_{0,{\rm max}}}$, 
which is quite a non-trivial constraint on many models of inflation. 
If this bound is saturated, the isocurvature perturbation may  
be observed in future CMB data.

\section{Predictions and Conclusions}
\label{sec:prdct-cncl}

In this paper we have made the observation that by promoting the $\mu$
parameter of the minimal supersymmetric standard model to a field, both
the strong CP and $\mu$ problems are solved.
How will we know whether this theory is correct? 

The first, and most important, test that
the theory must pass is that the three parameters $\mu$, $A$ and $\tan
\beta$, which are independent in the MSSM, must satisfy the relation
of equation (\ref{eq:mu}) \cite{MN}. This signals that the electroweak 
symmetry breaking sector of the theory is governed by 
the superpotential interaction $\lambda S H_1 H_2$, without 
the $S^3$ interaction of the next-to-MSSM, and that the vacuum
is the one with $\vev{s} = \tilde{m}/\lambda$ that occurs 
when $|m_S^2| < \lambda^2 v^2$ \cite{CP,HW}. 
While verification of this relation will show how the $\mu$ problem 
is solved, it is clearly insufficient to demonstrate that 
there is a PQ solution to the strong CP problem. 
For example, it could be that $\lambda \approx 10^{-3}$ and 
that small explicit symmetry breaking terms give the would-be axion 
a mass of order a GeV \cite{HW}, so that the strong CP problem is 
not solved. 
On the other hand,  observing a very small value for $\lambda$ 
would provide a strong indication that $\vev{s} = \tilde{m}/\lambda$ 
is large and is consistent with astrophysical constraints 
on the PQ solution. Such evidence for small $\lambda$ could be found
in the cascade decays of the superpartners produced at hadron
colliders, as we now discuss.

Recall that all the superpartners
have masses of order $\tilde{m}$, except for the axino and the
gravitino which are much lighter. An important question is the decay
mode and decay rate of the lightest superpartner amongst those that
have masses of order $\tilde{m}$, the LSP$'$. For large $\lambda$
the dominant decay of the LSP$'$ will be to the axino, $\tilde{a}$. 
For example, if the LSP$'$ is a neutralino, it would decay with a 
rate $\lambda^2 \tilde{m}$ to either $h \tilde{a}$ or $Z \tilde{a}$,
leading to the spectacular events discussed in \cite{HW}. 
The last decay process of the cascade chain takes place 
well inside the beam pipe. 
However, in axionic theories we have $\lambda \lsim 10^{-9}$ and 
a low value for $\sqrt{F_{\rm SUSY}}$ so that LSP$'$ decays 
to gravitinos, $\tilde{G}$, have a branching ratio comparable to 
or higher than that of decays to axinos.
The decay rate to the gravitino is $\tilde{m}^5 / F_{\rm SUSY}^2$, 
and hence 
\begin{equation}
\frac{\Gamma(LSP' \rightarrow \tilde{a})}
     {\Gamma(LSP' \rightarrow \tilde{G})}
\simeq \left( \frac{\lambda}{10^{-9}} \right)^2 
       \left( \frac{\sqrt{F_{\rm SUSY}}}{300 \;  \TEV}  \right)^4 
  \lsim 1.
\label{eq:lsp'decay} 
\end{equation}
When $\sqrt{F_{\rm SUSY}} \ll 300$ TeV, the LSP$'$ decays are mainly 
to gravitinos. 
The decay vertices are still within the beam pipe, and the event
configuration of decays to gravitinos is quite similar to that of 
decays to axinos. 
Thus, it may be difficult to distinguish this axionic theory 
with $\sqrt{F_{\rm SUSY}} \ll 300$ TeV from the theory  
with $\lambda \sim 10^{-3}$ in \cite{CP,HW}.
In cases with photino LSP$'$, however, 
${\rm Br}({\rm LSP}' \rightarrow \tilde{a}+\gamma)/
 {\rm Br}({\rm LSP}' \rightarrow \tilde{a}+Z)$ is different from 
${\rm Br}({\rm LSP}' \rightarrow \tilde{G}+\gamma)/
 {\rm Br}({\rm LSP}' \rightarrow \tilde{G}+Z)$. Determination of 
the mixing in the neutralino sector might be able to exclude one of the 
two possibilities above.
When $\sqrt{F_{\rm SUSY}}$ is close to its upper bound of 300~TeV, 
the LSP$'$ decay to a gravitino occurs at vertices separated from 
the primary vertex.\footnote{In the
  previous section we pointed out that one way to accomplish
  baryogenesis at low temperatures is by introducing large $R$ parity
  violating operators of the form $UDD$. In this case the LSP$'$ is
  expected to decay dominantly via these interactions, so that the
  separated vertex signal is replaced by the signals of $R$ parity  
  violation.} While this is also a typical signal of gauge-mediated 
SUSY breaking with such a value of $F_{\rm SUSY}$, decays at separated 
vertices, combined with a positive test of the relation (\ref{eq:mu}) 
for $\mu$, would give a strong indication that $\lambda$ is small, and that 
our proposal for simultaneous solutions of strong CP and $\mu$ problems 
is correct.

It may also happen that the QCD-charged extra particles introduced 
in section \ref{sec:thermal} are within the reach of hadron colliders. 

If our theory is correct, hadron colliders will tell us that 
the LSP is not in the MSSM sector. They will also tell us, 
from the LSP$'$ decay rate, that the gravitino mass is too small 
for gravitino cold dark matter.
The axion will be the natural remaining candidate for cold dark 
matter. 

When $f_a$ saturates the lower bound $10^{11}$ GeV, and 
$T_R \sim 1$~GeV, the cold dark matter consists of 
both saxions and axions.
The standard axion dark matter search \cite{haloscope} can 
detect the axion of this theory. 
The saxion can be detected indirectly from the soft X rays 
produced by its subdominant decay mode $s \rightarrow \gamma\gamma$.
There are two sources of this X-ray flux. One is from the saxions 
distributed uniformly throughout the universe, and the other arises
from saxions which have 
fallen into the gravitational potential of clusters of galaxies 
({\it c.f.} \cite{Turner2}).
The former is observed as an isotropic flux with a continuous 
spectrum, because the X rays emitted long ago and far away 
appear red-shifted. A rough estimate for the photon flux is given by 
\begin{eqnarray}
 \frac{d \Phi_{\rm isotropic}}{d\Omega \, dE_\gamma} 
   & \approx & \frac{3}{8 \pi} 
      \frac{n_s \Gamma(s \rightarrow \gamma \gamma)c t_0}{m_s c^2/2} 
      \sqrt{\frac{E_\gamma}{m_s c^2/2}}  \nonumber \\
   & \approx & 10^3  \; \sqrt{\frac{E_\gamma}{m_s c^2/2}} \; 
         {\rm /cm^2/sec/str/keV}  \nonumber \\
   & & \qquad \times 
     \left(\frac{100 \, \GEV}{T_{\rm PQ}}\right)^4
     \left(\frac{T_R}{1\, \GEV}\right)^5
     \left(\frac{m_s c^2}{100 \, \EV}\right) 
     \left(\frac{10^{11}\, \GEV}{f_a}\right)^2
\label{eq:unif-X}
\end{eqnarray}
for $E_\gamma < m_s/2$, where matter dominance is assumed in the first
line, and the number density of the saxions in (\ref{eq:ns}) is used 
in the second line. 
This is the flux predicted in extragalactic space --- the flux on Earth
is reduced by absorption in the Galaxy, and will have a modified energy 
spectrum. Using ROSAT data,
the observed extragalactic soft X-ray background is found to be 
$30\mbox{--}65 \; \KEV/{\rm cm}^2 /{\rm sec} /{\rm str} /\KEV$ for the
1/4 keV energy region \cite{1/4}.
Given the uncertainties in the predicted flux of (\ref{eq:unif-X}), 
the X-ray flux from saxion decays is consistent with observation.
We note here that the saxion mass $m_s$ could be as high as 1 keV
in a parameter region, although we have typically considered 
$m_s \approx 100$~eV.

Observations with high angular resolution suggest that 
60\% (and perhaps more) of the extragalactic X-rays can be attributed to 
the flux from discrete sources, such as AGN's, in the 1--2 keV energy region 
\cite{H93}. But it is not clear whether all of the soft X-ray 
flux is accounted for by such sources in the sub-keV energy region 
\cite{McS,FB,H93,1/4}, and there is still room for extra fluxes 
with particle-physics origins. 
Further observation with high angular resolution and long exposure 
time will certainly help determine the purely isotropic extragalactic 
component while removing foregound contamination and contributions from 
discrete sources. 
Furthermore, our X-ray signal from CDM saxions has 
a different spectrum from those of the discrete sources identified 
in \cite{H93}. Those fluxes decrease with increasing energy 
and have tails that extend above 1 keV \cite{H93}, while our 
signal's flux increases with energy until it is sharply 
cut off at an energy of half the saxion mass. 
Thus, observations with high energy resolution \cite{high-E-res}
will help identify the X-ray signal from CDM saxions, 
when combined with a better understanding of the foreground 
absorption and emission.  

The X-rays from saxions bound to a galactic
cluster produce a ``line spectrum'', with a width 
$\Delta E_\gamma / E_\gamma \sim {\cal O}(v/c) \sim 10^{-2}$, 
and an energy $m_s/(2(1+z))$, with $z$ the redshift of the cluster.
The photon flux at the peak energy is roughly given by 
\begin{eqnarray}
 \frac{d \Phi_{\rm cluster}}{d\Omega \, dE_\gamma} & \approx &
   \frac{1}{4\pi} \frac{1}{10^{-2}E_\gamma} \int n_s|_{\rm cluster} 
       \Gamma(s\rightarrow \gamma\gamma), \nonumber \\
  & \approx & 3 \times 10^4 \, 
          /{\rm cm}^2/{\rm sec}/{\rm str}/{\rm keV}
    \nonumber \\
  & & \quad \times  
      \left(\frac{\Omega_s h^2}{5 \Omega_B h^2}\right)
      \left(\frac{{\rm cluster~size}}{1 {\rm Mpc}}\right)
      \left(\frac{n_B|_{\rm cluster}}{10^{-3} /{\rm cm}^3}\right) 
      \left(\frac{m_s c^2}{100 \, \EV}\right)
      \left(\frac{10^{11}\, \GEV}{f_a}\right)^2,
\end{eqnarray}
where the integration in the first line is along the line of sight.
In order to identify these line-spectrum X rays, observations with 
both high angular resolution and high energy resolution are required.

As $T_R$ is lowered below 1 GeV,  $f_a$ increases above $10^{11}$ GeV 
and these axion and saxion signals become too difficult to see: 
the axion detection rate is proportional to  
$\lambda \approx 1/(m_a f_a^2)$ \cite{haloscope}, and 
the extragalactic photon flux from saxion decay
$\propto \lambda^{6.8}$, where eq.~(\ref{eq:fatd}) and 
$E_{\gamma} \propto m_s \approx \lambda v$ are used.
However, the lowest value of $f_a$ (and hence the largest of 
$\lambda \sim 10^{-9}$) is the most theoretically well-motivated. 
So far in this paper, the extremely small value for $\lambda$ 
has been put by hand, 
but it can be obtained naturally as $\sqrt{\tilde{m}/M_{\rm pl}} \sim
10^{-9}$, where $M_{\rm pl}$ is the Planck scale. 
Indeed, one can think of a theory with a superpotential 
\begin{equation}
 W = \frac{1}{M}S' S H_1 H_2 + \frac{1}{M'}S^{'4}, 
\label{eq:modelB}
\end{equation}
where $M \sim M' \sim M_{\rm pl}$. With a negative 
SUSY-breaking mass-squared of order $-\tilde{m}^2$ for $S'$, 
one can easily see that $\vev{S'}/M \sim \sqrt{\tilde{m}/M_{\rm pl}}$.
The mixing between $S$ and $S'$ is so small in this theory that 
the phenomenological analysis given in this article with an effective 
coupling $\lambda \equiv \vev{S'}/M$ is completely valid.

The theory with the superpotential (\ref{eq:model}) is quite similar 
to those \cite{SUSY-DFSZ} with 
\begin{equation}
 W=\frac{1}{M}S'SH_1 H_2 + \frac{1}{M'}S^n S^{' (4-n)} \qquad 
  (n \neq 0,4)
\label{eq:modelC}
\end{equation}
at first sight, but these two classes of theories are quite different. 
In our theory ($n=0$), $S'$ is neutral under the Peccei--Quinn 
symmetry, and the mixing between $S'$ and $S$ is quite small. 
Thus, the chiral multiplet $S$ is virtually the only one responsible 
for the spontaneous Peccei--Quinn symmetry breaking. This is one of the 
most important reasons why the axino, which is the LSP, 
is extremely light in our theory ({\it c.f.} \cite{CKL}). 
Another important aspect of our theory is that the stabilization of 
$s$ results only after the electroweak phase transition, so that 
the $\mu$ parameter is predicted in terms of $\tan \beta$ and the $A$ 
parameter. 
Therefore, our theory is not merely a particular case of the theories 
in \cite{SUSY-DFSZ}, but is essentially different. Indeed, the theory 
of the invisible axion presented in this article has several predictions that 
can be tested in the near future.

\section*{Acknowledgments}

This work was supported in part by the Director, Office of Science, Office 
of High Energy and Nuclear Physics, of the US Department of Energy under 
Contract DE-AC03-76SF00098 and DE-FG03-91ER-40676, and in part by the 
National Science Foundation under grant PHY-00-98840.
T.W. thanks participants of SUSY 2004 at Tsukuba, Japan, and 
the Miller Institute for the Basic Research in Science.


\begin{thebibliography}{99}

\bibitem{PQ} 
%
R.~D.~Peccei and H.~R.~Quinn,
Phys.\ Rev.\ Lett.\  {\bf 38}, 1440 (1977), 
Phys.\ Rev.\ D {\bf 16}, 1791 (1977).
%
\bibitem{WW} 
%
S.~Weinberg,
Phys.\ Rev.\ Lett.\  {\bf 40}, 223 (1978); 
%
F.~Wilczek,
Phys.\ Rev.\ Lett.\  {\bf 40}, 279 (1978).
%
%
%
\bibitem{DFSZ} 
%
M.~Dine, W.~Fischler and M.~Srednicki,
Phys.\ Lett.\ B {\bf 104}, 199 (1981);
%
A.~R.~Zhitnitsky,
Sov.\ J.\ Nucl.\ Phys.\  {\bf 31}, 260 (1980)
[Yad.\ Fiz.\  {\bf 31}, 497 (1980)].
%
\bibitem{SUSY-DFSZ}
%
H.~P.~Nilles and S.~Raby,
Nucl.\ Phys.\ B {\bf 198}, 102 (1982);
%
J.E. Kim and H.P. Nilles, 
Phys.\ Lett.\ B {\bf 138}, 150 (1984).
%
%
%
\bibitem{SUSY-DFSZ2}
%
H.~Murayama, H.~Suzuki and T.~Yanagida,
Phys.\ Lett.\ B {\bf 291}, 418 (1992).
%
\bibitem{SUSY-DFSZ3}
%
M.~Bastero-Gil and S.~F.~King,
Phys.\ Lett.\ B {\bf 423}, 27 (1998)
[arXiv:hep-ph/9709502].
%
\bibitem{MN} 
%
D.~J.~Miller and R.~Nevzorov,
arXiv:hep-ph/0309143.
%
\bibitem{CP} 
%
P.~Ciafaloni and A.~Pomarol,
Phys.\ Lett.\ B {\bf 404}, 83 (1997)
[arXiv:hep-ph/9702410].
%
\bibitem{HW}
%
L.~J.~Hall and T.~Watari,
arXiv:hep-ph/0405109.
%
\bibitem{Raffelt-book}
%
G.~Raffelt, ``{\it Stars as Laboratories for Fundamental Physics,}'' 
U.~Chicago Press, 1996.
%
\bibitem{MNZ}
%
D.~J.~Miller, R.~Nevzorov and P.~M.~Zerwas,
Nucl.\ Phys.\ B {\bf 681}, 3 (2004)
[arXiv:hep-ph/0304049].
%
%
%
\bibitem{GNS}
%
W.~D.~Goldberger, Y.~Nomura and D.~R.~Smith,
Phys.\ Rev.\ D {\bf 67}, 075021 (2003)
[arXiv:hep-ph/0209158].
%
\bibitem{KSY}
%
E.~D.~Stewart, M.~Kawasaki and T.~Yanagida,
Phys.\ Rev.\ D {\bf 54}, 6032 (1996)
[arXiv:hep-ph/9603324].
%
\bibitem{KT}
%
E.~Kolb and M.~Turner, ``{\it The Early Universe,}'' 
Addison-Wesley, 1990.
%
\bibitem{DH}
S.Dimopoulos and L.J. Hall, 
Phys.\ Lett.\ B {\bf 196}, 135 (1987).
%
\bibitem{Davis}
%
R.~L.~Davis,
Phys.\ Lett.\ B {\bf 180}, 225 (1986); 
%
R.~L.~Davis and E.~P.~S.~Shellard,
Nucl.\ Phys.\ B {\bf 324}, 167 (1989);
%
A.~Dabholkar and J.~M.~Quashnock,
Nucl.\ Phys.\ B {\bf 333}, 815 (1990).
%
\bibitem{HS}
%
D.~Harari and P.~Sikivie,
Phys.\ Lett.\ B {\bf 195}, 361 (1987);
%
C.~Hagmann and P.~Sikivie,
Nucl.\ Phys.\ B {\bf 363}, 247 (1991).
%
%
%
\bibitem{Turner}
%
M.~S.~Turner,
Phys.\ Rev.\ D {\bf 33}, 889 (1986).
%
\bibitem{KMY}
%
M.~Kawasaki, T.~Moroi and T.~Yanagida,
Phys.\ Lett.\ B {\bf 383}, 313 (1996)
[arXiv:hep-ph/9510461].
%
\bibitem{iso}
%
A.~D.~Linde,
Phys.\ Lett.\ B {\bf 259}, 38 (1991).
%
\bibitem{haloscope}
%
P.~Sikivie,
Phys.\ Rev.\ Lett.\  {\bf 51}, 1415 (1983)
[Erratum-ibid.\  {\bf 52}, 695 (1984)], 
%
Phys.\ Rev.\ D {\bf 32}, 2988 (1985)
[Erratum-ibid.\ D {\bf 36}, 974 (1987)]; 
%
R.~Bradley {\it et al.},
Rev.\ Mod.\ Phys.\  {\bf 75}, 777 (2003).
%
\bibitem{Turner2}
%
M.~S.~Turner,
Phys.\ Rev.\ Lett.\  {\bf 59}, 2489 (1987)
[Erratum-ibid.\  {\bf 60}, 1101 (1988)]; 
%
M.~A.~Bershady, M.~T.~Ressell and M.~S.~Turner,
Phys.\ Rev.\ Lett.\  {\bf 66}, 1398 (1991).
%
%
%
\bibitem{1/4}
%
W.~Cui, W.~T.~Sanders, D.~McCammon, S.~L.~Snowden and D.~S.~Womble,
arXiv:astro-ph/9604129.
%
\bibitem{H93}
%
G.~Hasinger, R.~Burg, R.~Giacconi, G.~Hartner, M.~Schmidt, J.~Trumper, 
and G.~Zamorani,
Astron. Astrophys. {\bf 275}, 1, 1993.
%
\bibitem{McS}
%
D.~McCammon and W.~Snaders, 
Annu. Rev. Astron. Astrophys. {\bf 28}, 657, 1990.
%
\bibitem{FB}
%
A.~Fabin and X.~Bracons,
Annu. Rev. Astron. Astrophys. {\bf 30}, 429,1992.
%
\bibitem{high-E-res}
%
D.~McCammon {\it et al.},
Astrophys.\ J.\  {\bf 576}, 188 (2002)
[arXiv:astro-ph/0205012].
%
\bibitem{CKL}
%
E.~J.~Chun and A.~Lukas,
Phys.\ Lett.\ B {\bf 357}, 43 (1995)
[arXiv:hep-ph/9503233].
%

\end{thebibliography}
\end{document}